\documentclass[aps,prl,preprint,showpacs,groupedaddress]{revtex4}
\usepackage{graphicx}

\begin{document}

\title{A Field-Induced Re-Entrant Novel Phase and a Ferroelectric-Magnetic
Order Coupling in HoMnO$_{3}$}
\author{B. Lorenz$^{1}$, A. P. Litvinchuk$^{1}$, M. M. Gospodinov$^{2}$, and
C. W. Chu$^{1,3,4}$}
\affiliation{$^{1}$Department of Physics and TCSAM, University of Houston, Houston, TX
77204-5002}
\affiliation{$^{2}$Institute of Solid State Physics,
Bulgarian Academy of Sciences, 1784 Sofia, Bulgaria}
\affiliation{$^{3}$Lawrence Berkeley National Laboratory, 1 Cyclotron Road, Berkeley, CA
94720}
\affiliation{$^{4}$Hong Kong University of Science and Technology, Hong Kong, China}
\date{\today }

\begin{abstract}
A re-entrant novel phase is observed in the hexagonal ferroelectric
HoMnO$_{3}$ in the presence of magnetic fields in the temperature range
defined by a plateau of the dielectric constant anomaly. The plateau evolves 
with fields from a narrow dielectric peak at the Mn-spin rotation transition 
at 32.8 K in zero field. The anomaly appears both as a function of temperature 
and as a function of magnetic field without detectable hysteresis. This is 
attributed to the indirect coupling between the ferroelectric (FE) and 
antiferromagnetic (AFM) orders, arising from an FE-AFM domain wall effect.
\end{abstract}

\maketitle











The search for and study of a coupling between the ferroelectric and
magnetic orders in ferroelectromagnets, such as the hexagonal manganites,
are of both fundamental and applied significance. This is because, by
symmetry arguments, such a direct coupling in manganites is highly unlikely.
However, it is not inconceivable that, through certain secondary indirect
interactions, such a coupling exists and results in a magneto-dielectric
effect with an interesting device potential in which the dielectric (magnetic)
properties can be modified by the onset of a magnetic (dielectric)
transition or the application of a magnetic (electric) field. We have
therefore examined the dielectric and magnetic properties of hexagonal HoMnO$%
_{3}$ single crystals as functions of temperature and magnetic field.
Indeed, in the presence of a magnetic field below 4.1 T, we have detected a
new re-entrant phase in HoMnO$_{3}$ below its zero-field Mn-spin
reorientation transition temperature of 32.8 K, indicative of a coupling
between the magnetic and ferroelectric orders and introducing an additional
dimension into the intriguing magnetic phase diagram of the compound.

The hexagonal manganites, ReMnO$_{3}$ (Re = Sc, Y, Er, Ho, Tm, Yb, Lu), have
attracted attention because of a rich variety of antiferromagnetic (AFM) as well
as ferroelectric (FE) phases with high FE Curie temperatures between 
590 and 1000 K. The AFM order of the Mn spins and/or the 
Re moments (e.g. Ho) is stabilized at lower temperature and coexists with the 
FE polarization. The Mn ions form a triangular lattice in which 
the moments are coupled antiferromagnetically by superexchange via the in-plane 
oxygen ions. This gives rise to spin-frustration effects and an AFM spin
arrangement with neighboring spins rotated by 120$^{\circ}$. The magnetic
structures have been investigated by neutron 
scattering and second harmonic generation optical experiments 
\cite{Sugie,Fiebig1,Fiebig2,Lonkai,Munoz}.

Hexagonal HoMnO$_{3}$ is FE below 830 K and its Mn-sublattices 
exhibit AFM order below $T_{N}\approx $ 76 K. A sharp Mn-spin-reorientation 
transition from $P\underline{6}_{3}\underline{c}m$ to $P\underline{6}_{3}c\underline{m}$ 
magnetic symmetry takes place at $T_{SR}\approx $ 33 K together with another Mn-spin-reorientation close to 4
K. In $P\underline{6}_{3}c\underline{m}$ symmetry 1/3 
of the Mn-spins are aligned with the crystallographic $a$-axis, 
whereas in the $P\underline{6}_{3}\underline{c}m$ phase the Mn-spins are 
rotated in-plane by 90$^\circ$. The compound possesses a very interesting 
re-entrant magnetic phase
diagram \cite{Fiebig2}. In addition to the Mn-spin ordering, neutron
scattering experiments suggested an AFM ordering of part of the Ho-spins 
with the principal spin direction along the $c$-axis
below or close to $T_{SR}$ \cite{Munoz,
Lonkai}. The AFM orders below $T_{N}$ clearly coexist with the
FE order. Although the direct coupling between the
in-plane staggered magnetization and the polarization is not allowed by
symmetry \cite{Birss}, the indirect coupling via lattice strain or
other effects may lead to anomalies in the dielectric
constant, $\varepsilon (T)$, in passing through the magnetic transitions. The first search for the magneto-dielectric
coupling detected a small anomaly of $\varepsilon $ at $T_{N}$ in YMnO$_{3}$ %
\cite{Huang}. Similar anomalies were later found in other hexagonal
manganites \cite{Iwata,Sugie,Katsufuji}. Very recently a correlation between
the FE and AFM domain walls was shown to exist in YMnO$_{3}$, providing
further evidence of a coupling between the two orders \cite{Fiebig3}.
Similar coupling is therefore expected to exist in HoMnO$_{3}$ and to give
rise to novel phenomena due to the complex magnetic phase diagram of the
compound.

Well-shaped plate-like single crystals of HoMnO$_{3}$ of 2.5 x 2.5 mm$^{2}$ in size and between 50 and 300 $\mu$m thick were prepared as described elsewhere \cite{Litvin}. Magnetization measurements were conducted using the MPMS (Quantum Design) with the field parallel and perpendicular to the $c$-axis. The capacitance of two crystals (70 and 240 $\mu$m thick) with gold pads sputtered onto two parallel faces was measured using the K-3330 and HP-4285A LCZ meters at frequencies between 10 kHz and 1 MHz. The PPMS (Quantum Design) was employed to control temperature and magnetic field up to 7 Tesla.

The $c$-axis dielectric constant $\varepsilon (T)$\ of HoMnO$_{3}$ 
at zero magnetic field is shown in Fig. 1. The small anomaly of $\varepsilon (T)$\ at $T_{N}$ = 76 K is in agreement with previous reports \cite{Sugie,Iwata}. Additionally, a sharp peak (width < 0.6 K, $\sim $ 5 \% of the base-$\varepsilon $) at 32.8 K is unambiguously detected for the first time. The peak position and its relative magnitude are independent of the measurement 
frequency, suggesting that the peak is related to a phase transition in the magnetic subsystem. The Mn-spin rotation transition was previously observed in HoMnO$_{3}$ at a $T_{SR}$ varying between 33 and 45 K, depending on the sample quality and the measurement techniques. The recent investigation of the magnetic phase diagram of single-crystalline HoMnO$_{3}$ \cite{Fiebig2} gives a $T_{SR}$ = 32.8 K, precisely the temperature at which the $\varepsilon $-peak is observed by us. We have therefore associated the $\varepsilon $-peak with the Mn-spin-reorientation transition.

In a magnetic field $H$ parallel to the $c$-axis the $\varepsilon $%
-anomaly broadens and evolves into a plateau-like structure with a sharp
increase at $T_{1}(H)$ and a quick drop at $T_{2}(H)$. The $\varepsilon (T)$%
's at different magnetic fields are shown in Fig. 2 with their
vertical axes shifted for clarity. While the overall $\varepsilon $%
-anomaly decreases and both $T_{1}$ and $T_{2}$ move toward lower
$T$ with $H$, the $\varepsilon (T)$ outside the peak is not
affected at all by $H$. The values of $\varepsilon (T)$ on
both sides of the peaks are the same for all data sets and coincide with
curve 5 in Fig. 2. At higher fields a second $\varepsilon $-plateau develops 
at lower T as demonstrated by the 3.3 Tesla data
(curve 3 in Fig. 2a). With further increasing $H$ the two
plateaus move toward one another and merge at 3.5 Tesla, forming one single
broad feature with a width of 15 K (curve 4 at 3.7 Tesla). The anomaly
disappears completely above 4.1 Tesla (curve 5). All
measurements were done with increasing as well as decreasing T and
no hysteresis was detected. The results are summarized as open
circles in Fig. 3, showing $T_{1}$ and $T_{2}$ as function of $H$. Both $T_{2}$ and $T_{1}$ exhibit re-entrance and disappear at fields
exceeding 3.5 and 4.1 T, respectively. They separate the novel phase
(I-phase) from the $P\underline{6}_{3}\underline{c}m$ and the $P\underline{6}%
_{3}c\underline{m}$ phases. The $\varepsilon $-value of the I-phase, characterized by the $\varepsilon $-plateau, decreases smoothly with
T as $(T_{c}-T)^{-n}$ with $n=0.01$ and $T_{c}=32.8\ K$ (see inset of Fig. 2). The $T_{1}(H)$ and $T_{2}(H)$ curves are similar to the phase boundary between the AFM (B$_{1}$) and the AFM (B$_{2}$) in Ref. \cite{Fiebig2} above 8 K, the present experimental limit for our dielectric measurements. The reentrant field in Ref. \cite{Fiebig2} is $\sim $ 3.8 T,
falling in between the 3.5 and 4.1 T observed by us. This makes it difficult
to assign the phase boundary in Ref. \cite{Fiebig2} to $T_{1}$ or $T_{2}$.

To verify the data of Fig. 3 and to support our conclusion about
the existence of an intermediate phase, we have conducted isothermal
measurements of $\varepsilon (H)$ that should show similar anomalies as the phase boundaries are crossed. $\varepsilon (H)$ is displayed in Fig. 4 for three temperatures. Each isothermal $\varepsilon (H)$ shows the expected enhancement in the I-phase with only one pronounced plateau-like structure. The $H$ dependence of $\varepsilon $ within each phase ($P\underline{6}_{3}\underline{c}m$, I-phase, and  $P\underline{6}_{3}c\underline{m}$) is weak, as already suggested from the data
of Fig. 2. The two transitions at $H_{2}$ (low field edge) and $H_{1}$ (high field edge) are included as solid circles in Fig. 3 and they are in perfect agreement with the results obtained from $\varepsilon (T)$ at constant $H$. The data show no hysteresis at the transitions with increasing and decreasing $H$.

In an attempt to determine the origin of the dielectric anomaly associated
with the I-phase, we examined the magnetic ordering of the Ho- and Mn-ions.
An AFM order of the Ho-spins along the $c$-axis was detected in neutron scattering experiments below 25 K \cite{Munoz}, as well as below 32.5 K \cite{Lonkai}. The details of this AFM order of the Ho spins are not resolved yet. Ref. \cite{Munoz} suggested that only 2/3 of the Ho moments participate in this ordering, leaving 1/3 of Ho spins disordered with a large paramagnetic contribution to the dc susceptibility, as observed in magnetic measurements \cite{Munoz,Sugie}. It is also not clear how the Ho-spin ordering is correlated with the Mn-spin rotation transition that stretches between 32.5 and 42 K in the 
neutron scattering studies of polycrystalline samples \cite{Lonkai}, in contrast to the results from single crystals in which the spin rotation transition is 
very sharp in zero field and appears close to 32.8 K \cite{Fiebig2}. Ho-spin ordering is expected to affect the $c$-axis magnetic susceptibility, $\chi _{c}$. We have therefore conducted magnetization measurements on our HoMnO$_{3}$ single crystals with the external magnetic field parallel and perpendicular
to the $c$-axis. Indeed, for the first time, we found a small but distinct
leveling-off of $\chi _{c}$ over a narrow temperature range (< 1 K), as 
indicated by the arrow in Fig. 5 (enlarged in the lower inset). The
anomaly is very sharp (as shown by $d\chi _{c}/dT$, upper inset in Fig. 5) and appears at exactly the same temperature (32.8 K) as the peak of $\varepsilon (T)$ at zero field. When an external field is applied, $d\chi _{c}/dT$ reveals additional anomalies that define $T_{1}$ and $T_{2}$, similar to $\varepsilon $. The transition temperatures so obtained are also shown in Fig. 3 as solid stars. No anomaly was observed in the in-plane ($H$ $\bot $ $c$-axis) susceptibility at these temperatures. Therefore, we conclude that the $\chi _{c}$ signals the onset of the AFM ordering of Ho moments oriented along the $c$-axis and the corresponding transition temperature coincides with the $T_{SR}$ of the Mn$^{3+}$. The two magnetic subsystems (Ho$^{3+}$ and Mn$^{3+}$) are obviously correlated and one may speculate that the onset of the Ho spin ordering triggers the spin rotation of the Mn moments.

A direct coupling between the polarization and the in-plane magnetization of
the Mn sublattice in both phases, $P\underline{6}_{3}\underline{c}m$ and $P%
\underline{6}_{3}c\underline{m}$, is not allowed by symmetry \cite{Birss}.
For the I-phase we tentatively assume that its magnetic structure is characterized by a Mn-spin order in which the moments rotate to an angle between 0 and 90$^\circ $ with respect to the $a$-axis. The resulting magnetic symmetry is $P\underline{6}_{3}$. This symmetry has been observed over a wider temperature range in LuMnO$_{3}$ and ScMnO$_{3}$ in zero magnetic field \cite{Fiebig1}. However, even in the reduced symmetry $P\underline{6}_{3}$
a direct coupling between the polarization and an in-plane magnetic moment is not allowed. The only magnetic space groups that can account for this type of coupling are $\underline{2}$, $m$, $\underline{2}/m$, $1$, and $\underline{1}$ \cite{Birss}\, but until now there is no evidence for such a low symmetry from  previous reports. The situation changes if the Mn moments tilt out-of-plane with a component along the $c$-axis, which is allowed in $P\underline{6}_{3}c\underline{m}$ (low-T phase) and $P\underline{6}_{3}$ (I-phase), but not in $P\underline{6}_{3}\underline{c}m$ (high-T phase). The $c$-axis magnetic moment can couple directly to the polarization. Our observations can be explained if we assume that the Mn magnetic moment tilts out-of-plane in the I-phase ($P\underline{6}_{3}$) and that canting toward the $c$-axis causes the enhancement of the dielectric constant via a $z-z$ coupling of magnetization and polarization. However, no definite conclusion for or against this possible scenario can be drawn at this time.

An alternative explanation of the magneto-electric coupling leading to dielectric anomalies as observed in our experiments is related to the physics of domain walls. Although our sample is ferroelectrically ordered, it is not a single domain. FE domain walls play a crucial role in the dielectric properties at low T. $\varepsilon$ is affected by the electric-field-induced movement of the FE domain walls. It was recently proposed \cite{Sakano} and experimentally proven \cite{Fiebig3} that in the low-$T$ $P\underline{6}_{3}c\underline{m}$ magnetic phase of YMnO$_{3}$ the FE domain wall forces the magnetic order parameter to change sign, i.e. each FE domain wall coincides with an AFM domain wall (although magnetic domain walls may exist solely within a single FE domain). An unambiguous explanation for this unusual effect is not yet available, but it has been speculated that the lattice strain in a FE domain wall couples to the Mn magnetic moment resulting in a decrease of free energy whenever the magnetic order parameter changes sign across the FE domain wall \cite{Fiebig3}. The "pinning" of the AFM domain wall to the FE wall will reduce the dielectric constant since the external electric field shifting the FE domain wall also has to move the correlated magnetic domain wall and this costs extra energy. In fact, all hexagonal rare-earth manganites show a slight decrease of $\varepsilon (T)$ in passing through $T_{N}$ into the AFM phase. If we assume that this pinning is effective in both high symmetry ($P\underline{6}_{3}\underline{c}m$ and 
$P\underline{6}_{3}c\underline{m}$) magnetic phases of HoMnO$_{3}$, but not in the intermediate phase the relative enhancement of $\varepsilon (T)$ in the I-phase as observed in our experiments could be explained. In this way the dielectric anomalies are caused by a domain wall pinning effect and $\varepsilon$ serves as a sensitive probe to distinguish between the different magnetic symmetries and phases in HoMnO$_{3}$. More experimental and theoretical work is needed, however, to unambiguously identify the physical mechanisms leading to the interesting observations reported here. In particular, calculations of the domain wall contributions to the free energy based on microscopic models including the magnetic superexchange interactions as well as the lattice distortions due to the FE order could lead to a better understanding of the magneto-electric coupling in the rare-earth hexagonal manganites.


\begin{acknowledgments}
The authors thank M. Iliev for stimulating discussions. This
work is supported in part by NSF Grant No. DMR-9804325, the T.L.L. Temple
Foundation, the John J. and Rebecca Moores Endowment, and the State of Texas
through TCSAM and at Lawrence Berkeley Laboratory by the Director, Office of 
Science, Office of Basic Energy Sciences, Division of Materials Sciences and 
Engineering of the U.S. Department of Energy under Contract No. DE-AC03-76SF00098. 
The work of M. M. G. is supported by the Bulgarian Science Fund, Grant No. F 1207.
\end{acknowledgments}


\begin{figure}
\caption{Low temperature dielectric constant of HoMnO$_3$ showing two
anomalies at the onset of magnetic order ($T_{N}$) and the spin rotation
transition ($T_{SR}$). Inset: details of the peak at $T_{SR}$.}
\label{Fig1}

\caption{$\protect\varepsilon (T)$ for selected external magnetic fields 
{\textit H}. (1) {\textit H}=0, (2) {\textit H}=2.6 Tesla, (3) {\textit H}=3.3 Tesla, 
(4) {\textit H}=3.7 Tesla, (5) {\textit H}=4.1 Tesla. 
Different curves are offset by a constant (indicated by dotted lines). 
$T_{1}$ and $T_{2}$ are marked by vertical bars next to curve 3. 
Inset: all data plotted on the same scale.}
\label{Fig2}

\caption{Low-temperature magnetic phase diagram of HoMnO$_{3}$. Open 
circles: data from $\protect\varepsilon (T)$ scans. Solid circles: data 
from isothermal field scans, $\protect\varepsilon (H)$. Solid stars: 
anomalies of the {\textit c}-axis magnetic susceptibility. Dotted line: 
phase boundary according to Ref. 5.}
\label{Fig3}

\caption{$\protect\varepsilon (H)$ for three selected temperatures. For 
the 26 K data the two transitions at $H_{2}$ and $H_{1}$ are indicated by vertical arrows.}
\label{Fig4}

\caption{{\textit c}-axis magnetic susceptibility of HoMnO$_{3}$. The 
arrow shows the anomaly associated with the AFM ordering of the Ho spins. 
Lower left inset: enlargement of the critical range. The derivative 
(upper right inset) shows a sharp peak at $T_{SR}$.}
\label{Fig5}
\end{figure}

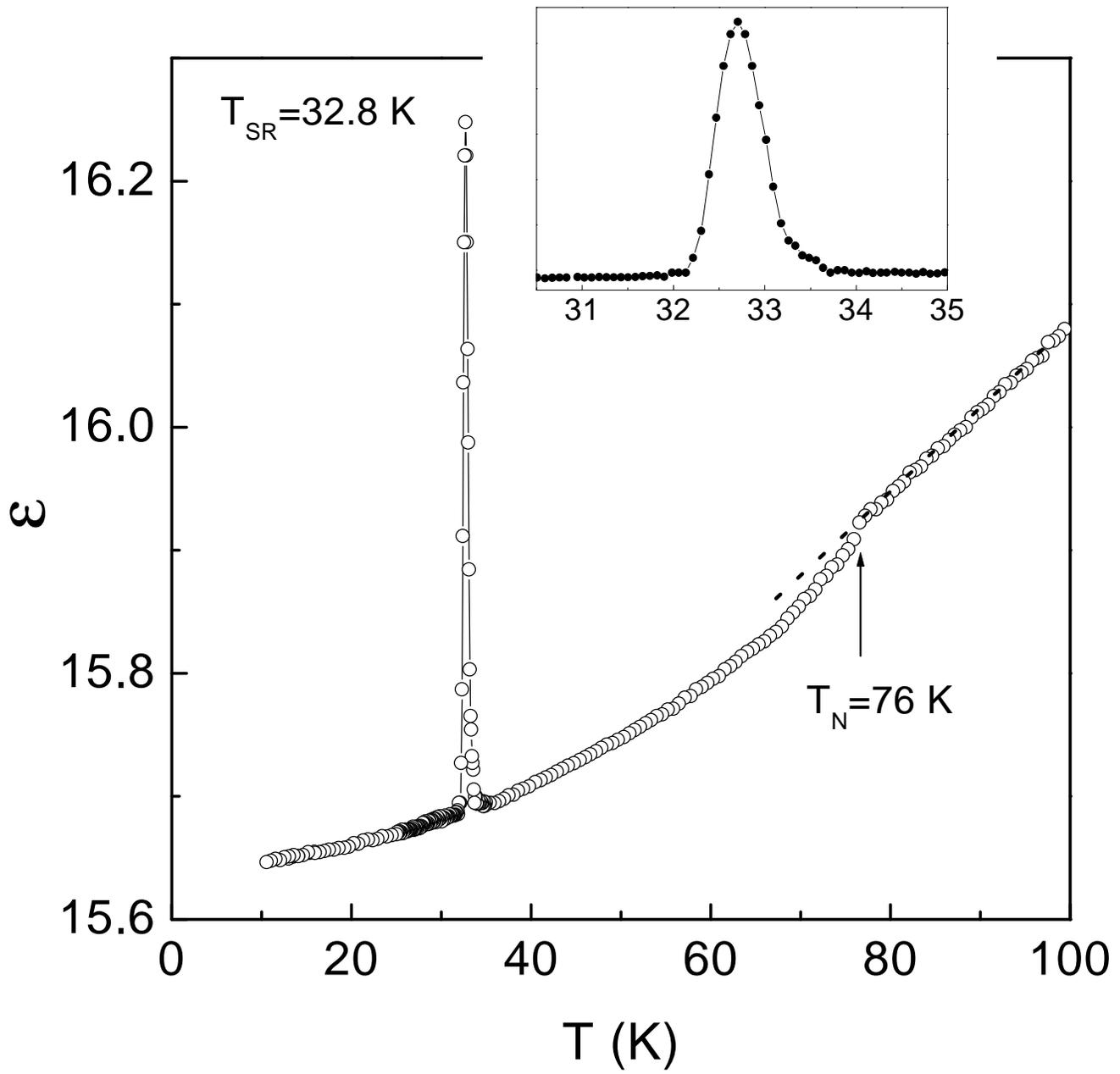
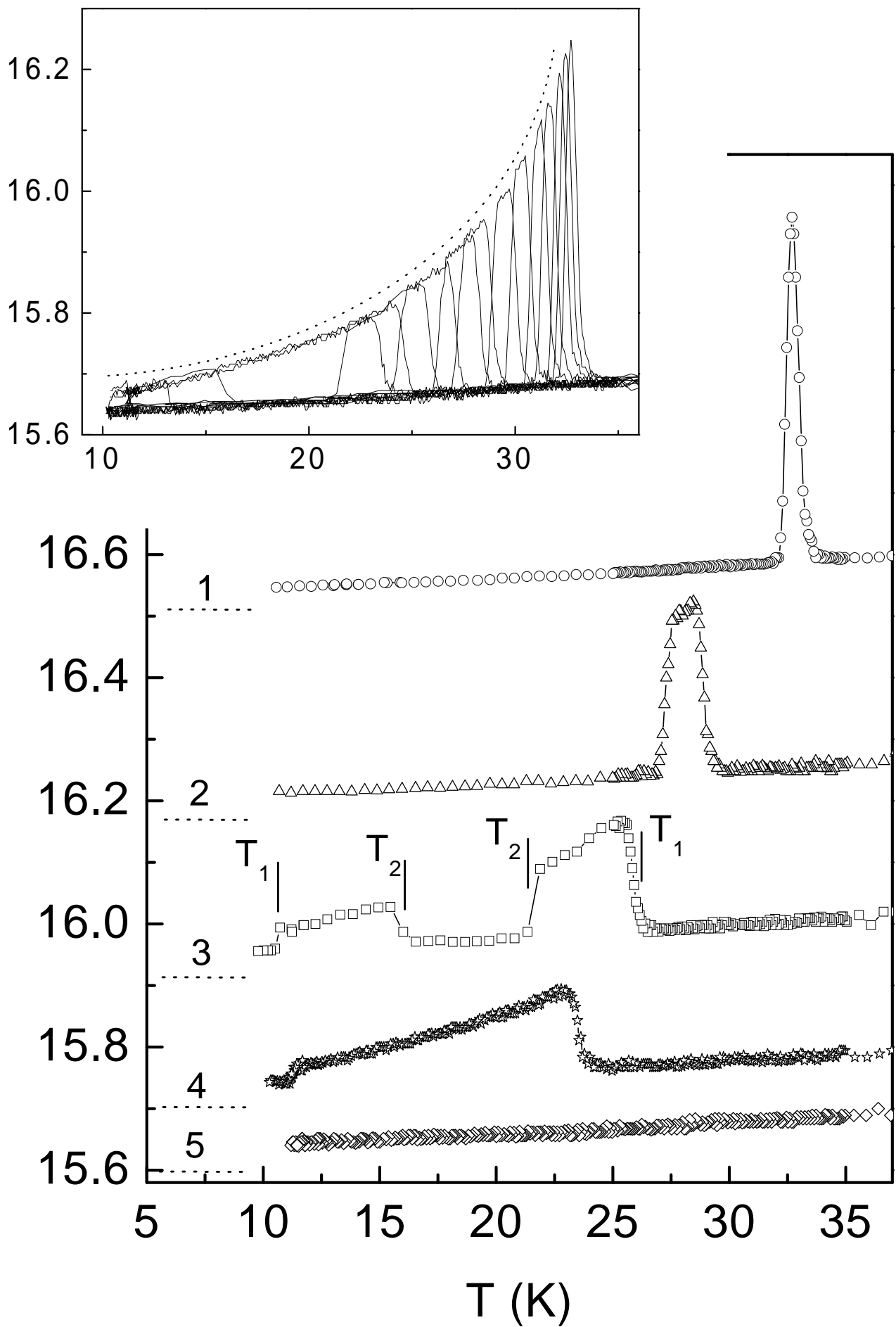
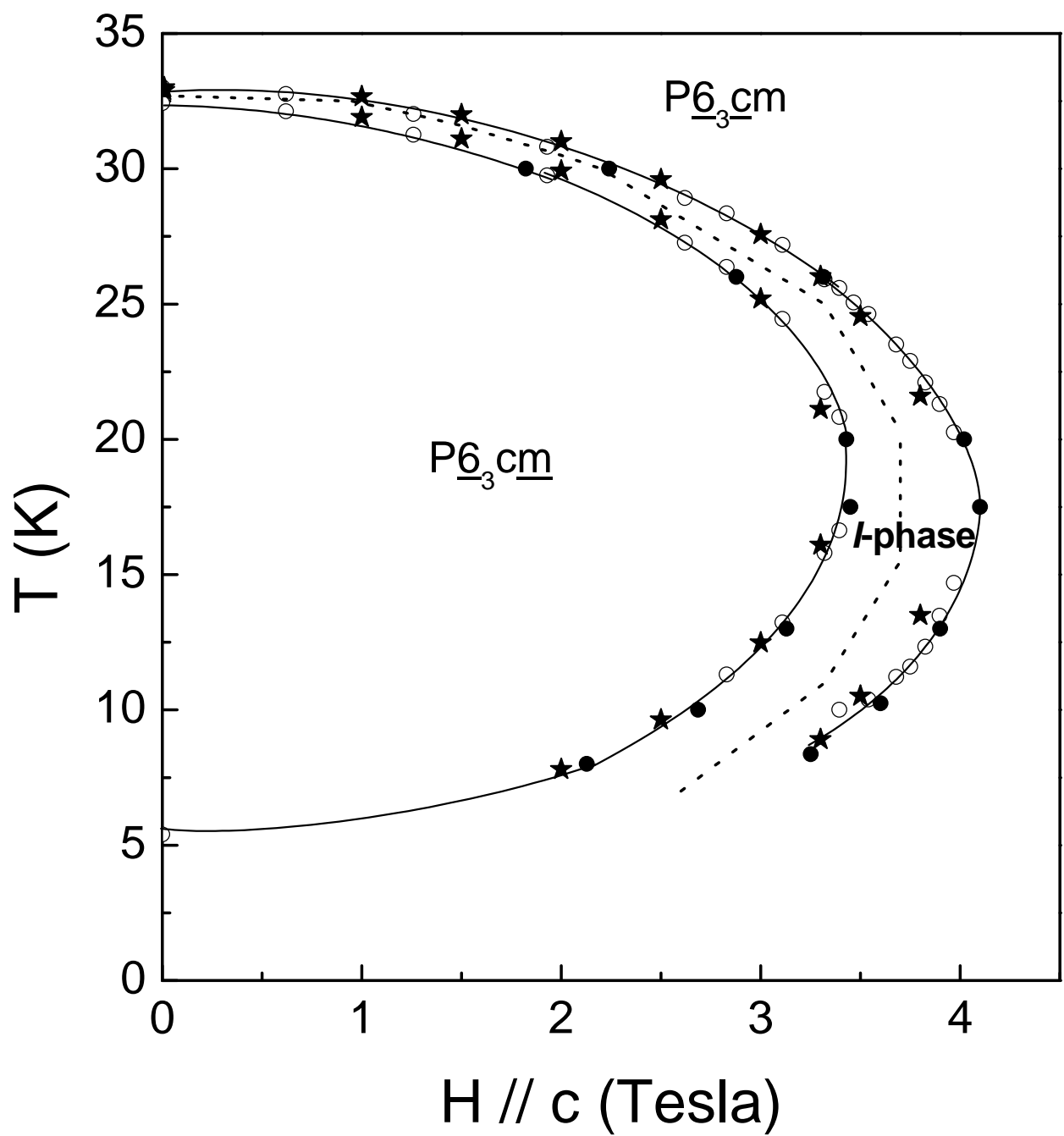
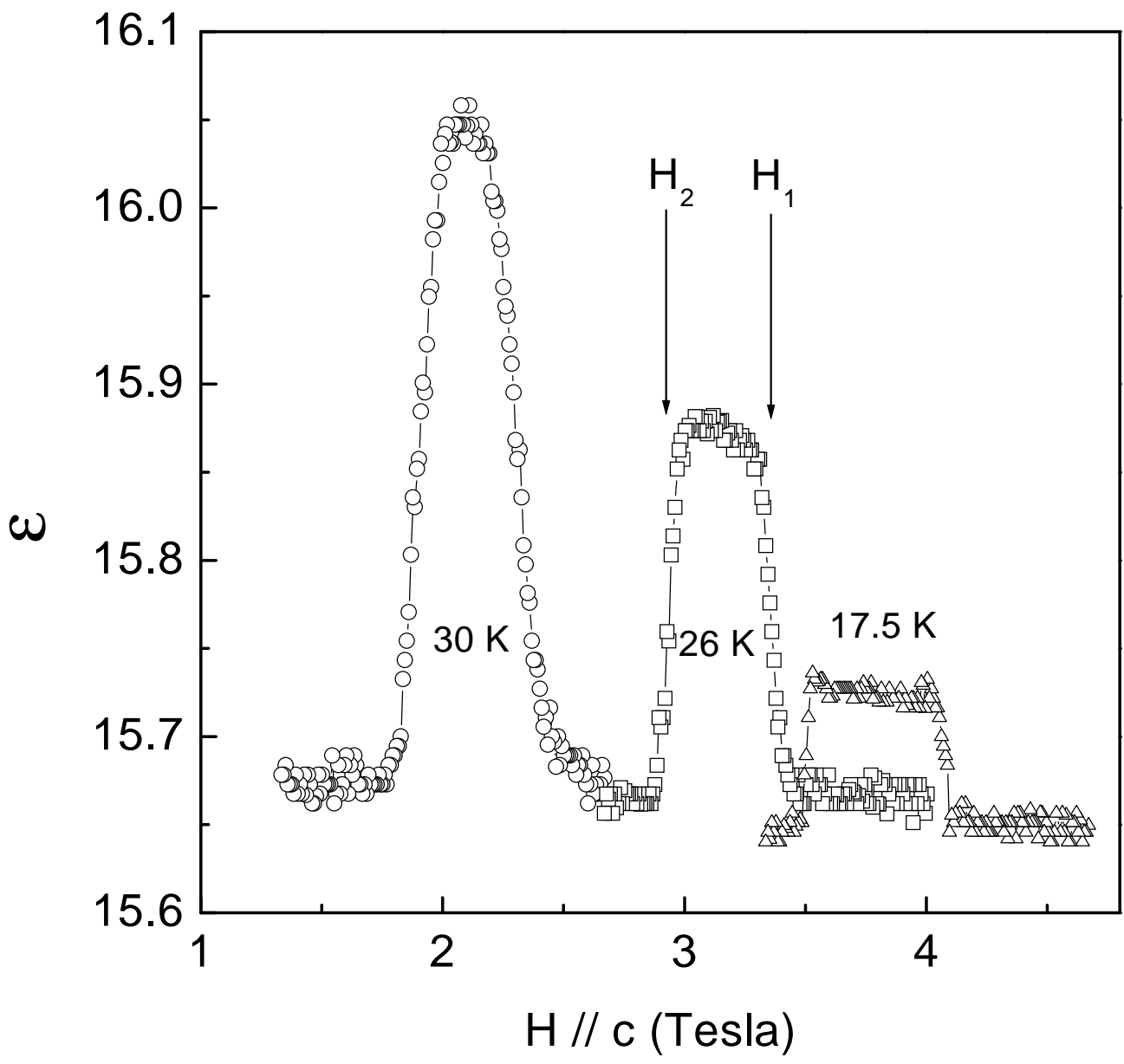
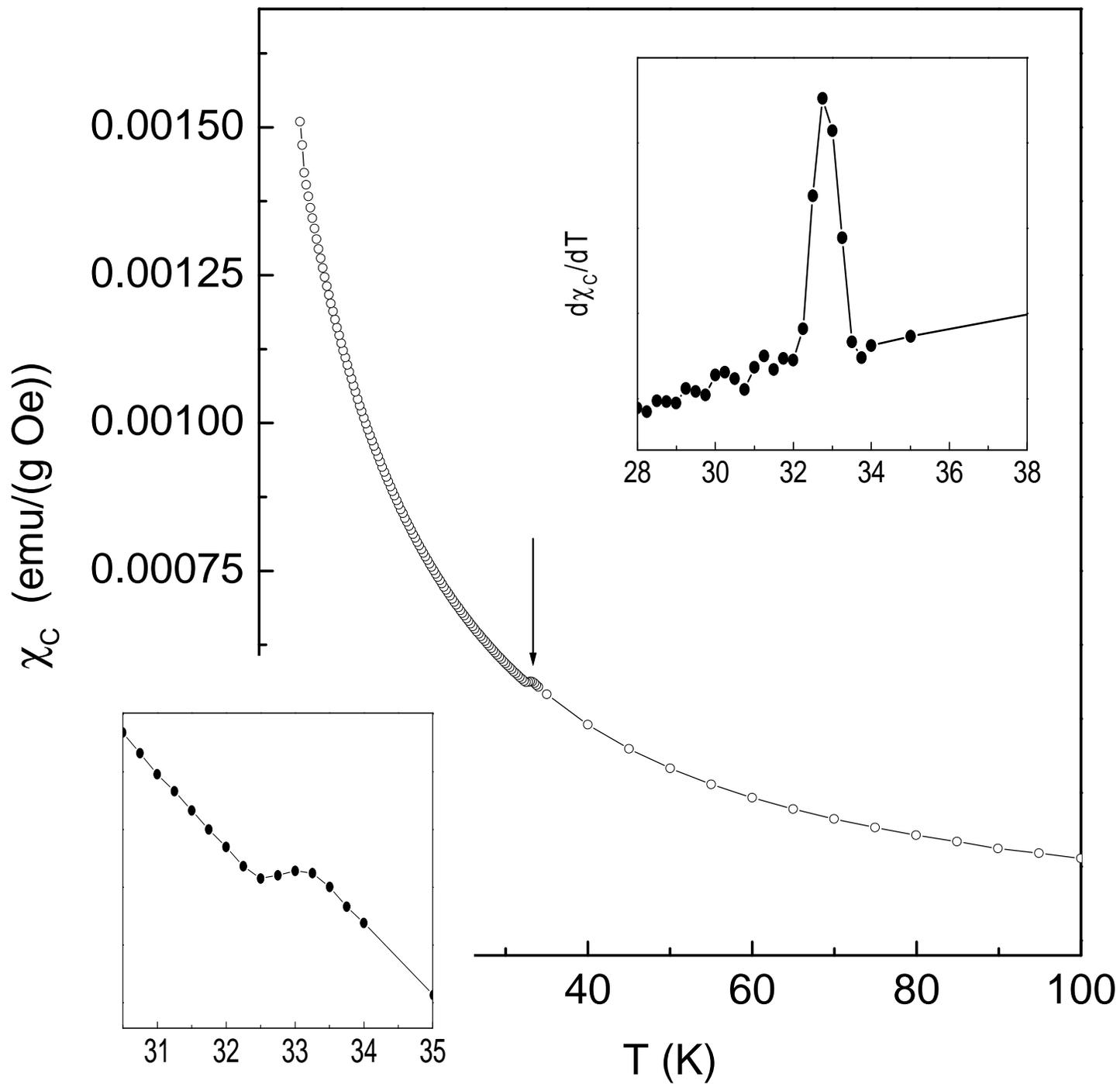
\end{document}